\documentclass[a4paper,10pt,showpacs,aps,floats,superscriptaddress]{revtex4}

\usepackage{epsfig}

\begin{document}

\title{Efficiency and reliability of epidemic data dissemination in
complex networks}
\author{Yamir Moreno}
\affiliation{Departamento de F\'{\i}sica Te\'orica, Universidad de
Zaragoza, Zaragoza 50009, Spain.}
\affiliation{Instituto de Biocomputaci\'on y F\'{\i}sica de Sistemas
Complejos, Universidad de Zaragoza, Zaragoza 50009, Spain}
\author{Maziar Nekovee}
\affiliation{Complexity Research Group, Polaris 134, BT Exact,
Martlesham, Suffolk IP5 3RE, UK}
\author{Alessandro Vespignani}
\affiliation{ Laboratoire de Physique Th\'eorique (UMR du CNRS 8627),
B\^atiment 210 Universit\'e de Paris-Sud 91405 ORSAY Cedex - France}

\date{\today}

\widetext
\begin{abstract}

We study the dynamics of epidemic spreading processes aimed at
spontaneous dissemination of information updates in populations with
complex connectivity patterns.  The influence of the topological
structure of the network in these processes is studied by analyzing
the behavior of several global parameters such as reliability,
efficiency and load. Large scale numerical simulations of update
spreading processes show that while networks with homogeneous 
connectivity patterns permit a higher reliability, scale-free
topologies allow for a better efficiency.
\end{abstract}
 

\maketitle

Modern society increasingly relies on large scale computer and
communication networks, such as the Internet.  A major challenge in
these networks is the developing of reliable algorithms for the  
dissemination of information from a given source to thousands, or even
millions, of users such as for news and stock exchange
updates, mass file transfers and Internet broadcasts \cite{dave,deering}. In
epidemic-inspired communication this is achieved by exploring a
mechanism analogous to the spreading of infectious diseases in
populations \cite{vogels,ep3}. Indeed,
epidemic data dissemination in computer and communication networks
show interesting parallels with both disease propagation in
populations and the spread of rumor in social networks.
The information spreads like a benign
epidemic through local interaction between nodes which forward the
message they receive to a random selection of their peers in the
network, until the whole system becomes ``infected'' with information.
The great advantages of epidemic-style communication is that
dissemination proceeds on a local basis, without any co-ordination
from a central organizing body \cite{vogels,ep3}. These 
protocols are also highly resilient to sudden failure of communication 
links and nodes.

A relevant result in the mathematical theory of epidemics is that the
spreading of infection in a population is strongly affected by the
patterns of connectivity in the underlying contact networks.  In
particular in scale-free topologies, characterized by degree
distributions with power law behavior $P(k)\sim k^{-\gamma}$
\cite{bar99,fal,romu01}, the statistical relevance of hubs makes the
network highly permeable to attacks \cite{newman00,havlin00,bar00} and
the spreading of infections \cite{pv01a} and highlights the need of
special immunization strategies. This result suggests that the
topology of the underlying computer and communication network might
heavily affect the performance of epidemic-style data dissemination
protocols. Surprisingly, however, the impact of network topology on
such protocols has not been thoroughly explored, although the results could
have an important technological value. Indeed, these protocols
potentially find a large spectrum of application such as mobile
communication networks and, more recently, resource discovery in the
so-called peer-to-peer systems built on top of the Internet
\cite{p2p}, and finally in grid computing \cite{grid}.

In this paper we define a simple epidemic data dissemination model and
perform a detailed numerical study of the
dynamics of the information propagation in networks with diverse 
topological properties. 
Our basic model is a slightly modified version of the Daley
and Kendall (DK) model \cite{dk64,zanette,liu} and it can be
considered as the simplest epidemic algorithm for the updating of
distributed databases \cite{ep1,ep3}. We study the main relevant
features of the model such as the reliability of the dissemination
process and the amount of traffic generated by the
dynamics. Specifically, we study two different prototypical networks:
a random homogeneous network~\cite{note1} and a scale-free network.
The results obtained point out that in the homogeneous topology the
epidemic process provides a more reliable updating of the network.
The scale-free topology, on the other hand, allows the algorithm to
perform more  efficiently in terms of the generated traffic. Finally, we 
compare the present model with a deterministic broadcast process and 
find that in a wide range of the model parameters the epidemic
algorithm is more efficient. 
  
The model we shall consider is defined in the following way.  Each of
the $N$ elements of the network can be in three possible
states. We call a node
holding an update and willing to transmit it a {\em spreader}. Nodes
that are unaware of the update will be called {\em ignorants} while
those that already know it but are not willing to spread the update
anymore are called {\em stiflers}.  We denote the density of
ignorants, spreaders and stiflers at time $t$ as $\psi(t)$, $\phi(t)$
and $s(t)$, respectively such that for all $t$,
$\psi(t)+\phi(t)+s(t)=1$. The spreading process takes place along the
links between spreaders and ignorants. Each time step spreaders contact one
(or more) neighboring node. When the
spreader contacts an ignorant, the last one turns into a new spreader
at a rate $\lambda$. On the other hand, the spreader becomes a stifler
with rate $1/\alpha$ if a contact with another spreader or a
stifler takes place~\cite{note2}. The parameter $\alpha$ can be
considered as the average number of contacts with spreader/stifler
nodes before the spreader turns itself into a stifler.
This dynamics mimics the attempt of diffusing an
update or rumor by nodes which have been recently updated. At the same
time, if a node attempts too many times to communicate the update to
nodes which have already received it, it stops the process turning
itself into a stifler.  In other words, the node realizes that the
update has lost its novelty and becomes uninterested in diffusing it.
The present dynamics thus introduces a trade-off in maximizing the
number of updated nodes and minimizing the number of contacts
attempted. Obviously, the efficiency of the spreading process will
depend on the rate at which individuals loose interest in further
spreading of the rumor and the topology of the underlying network.

At the mean-field level and using the homogeneous assumption for the
network connectivity pattern, the time evolution of ignorants,
spreaders and stiflers is described by the simple set of equations
\begin{equation}
  \partial_t\psi(t)  = -\lambda\psi(t)\phi(t), 
\end{equation}
\begin{equation}
  \partial_t\phi(t) = +\lambda\psi(t)\phi(t)
  -\frac{1}{\alpha}\phi(t)(\phi(t)+s(t)),
\label{eq1}
\end{equation}
where $s(t)$ is obtained by the normalization
condition $s(t)=1-\psi(t)-\phi(t)$ \cite{dk64,zanette}.  The 
dynamics asymptotically evolves to
the state $\phi(\infty)=0$ in which the system is frozen. Noticeably,
in random homogeneous networks, the density $s(\infty)$ of elements
which are aware of the update is always a finite fraction of
the whole population \cite{dk64,zanette}.  The homogeneous assumption
is, however, not valid anymore in the case of heterogeneous scale-free
networks where it is known that spreading processes may show very
different properties \cite{pv01a}. In particular, an explicit dependence on
the nodes' degree $k$ must be included in the rate  equations. 
While a general analytical solution
cannot be obtained in this case, numerical studies on scale-free
networks can be used to evaluate the reliability and efficiency of
this process in more complex topologies\cite{zanette,liu}.

In the present investigation we used two specific network models.
First we consider the Barab\'asi-Albert (BA) network \cite{bar99}. In
this model, starting from a set of $m_0$ nodes, one preferentially
attaches each time step a newly introduced node to $m$ older
nodes. The procedure is repeated many times and a network with a power
law degree distribution $P(k)\sim k^{-\gamma}$ with $\gamma=3$ and
average connectivity $\langle k \rangle=2m$ builds up. This network is
a clear example of highly heterogenous network, in that the degree
distribution has unbounded fluctuations. As a reference of homogeneous
networks we considered the Watts-Strogatz (WS) network \cite{ws98} in
the case of complete random rewiring. In this case, one starts from a
ring with $N$ nodes, each of them connected symmetrically to $2K$
neighbors. With probability $p$ each link connected to a clockwise
neighbor is rewired to a randomly chosen node; otherwise it is
preserved. After enough iterations a random network with an
exponential connectivity decay for large $k$ and $\langle k
\rangle=2K$ is generated. Henceforth, we will use $m_0=m=3$ for the BA
network and $p=1$ and $K=3$ for the WS model giving 
$\langle k \rangle =6$ for both networks.

We have performed large scale numerical simulations by applying
repeatedly the rules stated above on BA and WS networks. Initially,
$\psi(0)=\frac{N-1}{N}$, $\phi(0)=\frac{1}{N}$, and $s(0)=0$, i.e., we
start from a single spreader who is willing to spread the update
through the network. At every time step, each of the $\phi N$
spreaders contacts all its neighbors in a random sequence, 
unless during a contact it turns into a stifler. In this case it immediately
stops contacting further nodes. This account for the larger transmission
capabilities of high degree nodes that can reach a larger number of 
neighbors as specified by the heterogeneous network topology. 
The dynamical rules of the model are applied in
parallel. The sizes of the networks used in the simulations carried
out range from $N= 10^3$ nodes to $N=10^5$ nodes and all numerical
results have been obtained by averaging at least over 10 different
networks and $10^3$ iterations. The parameter
$\lambda$ may be varying as for the case of communication networks
where it is known that the rate of packet loss is not always
zero. Nevertheless, without loss of generality, 
$\lambda=1$ since it just fixes the time scale by rescaling opportunely
the rate $1/\alpha$ in the Eq.~\ref{eq1}. On the
other hand, we vary the rate at which spreaders decide not to
communicate the update any more from $\alpha=1$ to $\alpha=10$ and
monitor several quantities of interest.

\begin{table}[t]
%\begin{ruledtabular}
\begin{tabular}{|c|c|c|c|c|c|c|c|c|}
\hline
$\alpha$ & 1 & 2 & 3 & 4& 5 & 6& 8& 9\\
\hline
$R_{WS}$ & 0.831& 0.962 & 0.986 & 0.994& 0.996 & 0.996 & 0.998 & 0.999\\
$R_{BA}$ & 0.368 & 0.684 & 0.781& 0.874  & 0.932 & 0.952 & 0.977 &0.987\\
\hline
\hline
\end{tabular}
\caption{Reliability of the epidemic  process, defined as the density of
nodes that has received the update, in the  WS and BA networks for
different values of the parameter $\alpha$. }
\label{tab1}
%\end{ruledtabular}
\end{table}

In order to characterize the propagation process we first focus on the 
{\em reliability} $R$ of the rumor propagation defined as the final 
density $s(\infty)$ of nodes that have got the update when the process 
dies out. For obvious practical purposes,
any algorithm or process that emulates an efficient spreading of a
given message or data packet will try to raise as much as possible
this magnitude. 
\begin{figure}[t]
\begin{center}
\epsfig{file=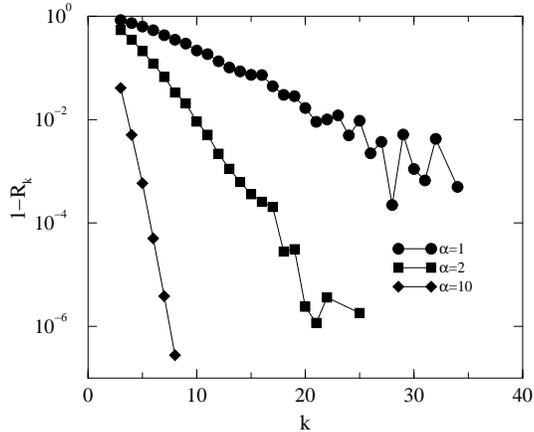,width=2.3in,clip=1,angle=-90}
\end{center}
\caption{Relative density of ignorants $1-R_k$ as a function of their
connectivity $k$ at the end of the spreading process in SF
networks for different values of the parameter $\alpha$ 
The size of the network is always $N=10^4$ nodes.}
\label{figure3}
\end{figure}
In Table~\ref{tab1} we report the reliability of the spreading process
for the BA network and the random graph generated with the WS
algorithm with $p=1$ for several values of the parameter $\alpha$. As
noticed previously \cite{liu}, in the WS network the number of stiflers
at the end of the process is already high even for $\alpha=1$, and the
BA network appears less reliable. In general, it results that
homogeneous networks allows a larger reliability $R$ to epidemic updating
processes.  This is not straightforward and one may think that the
existence of hubs in scale-free networks helps propagate the
rumor. However, a closer look at the spreading dynamics tells us that
the presence of hubs introduces conflicting effects in the dynamics.
While hubs may in principle reach a larger number of nodes,
spreader-spreader and spreader-stifler interactions get favored on the
long run.  Indeed, it is very unlikely that a hub in the spreader
state contacts all its ignorant neighbors before turning into a
stifler.  Once a few hubs are turned into stiflers many of the
neighboring nodes could be isolated and never get the update. In this
sense, homogeneous networks allow for a more capillary diffusion of
the update, since all nodes contribute equally to the message-passing.
This is opposite to what happens in the usual epidemic spreading model
in heterogeneous networks models. These models lack of an infected 
recovery rate induced by neighbors already infected and fully exploit the
advantage of hub's large degree~\cite{pv01a}.

The previous discussion, however, refers only to average properties.  In SF
networks the connectivity distribution is highly
heterogeneous and it is interesting to have a more detailed insight on
the reliability of the process for different connectivity classes. It
may be particularly relevant that a higher $R$ corresponds to the
highly connected nodes, the hubs, which have a dominant role in the
system. Figure\ \ref{figure3} shows the behavior of the reliability 
$R_k$ measured as a function of the nodes connectivity $k$. 
This amounts to the relative density of 
nodes with connectivity $k$ that have received the update and it is measured
as  $R_k=\langle\frac{S_k}{N_k}\rangle$, where $S_k$ and $N_k$ denote
the total number of stiflers and the total number of nodes with
degree $k$, respectively. The $\langle \rangle$ represents the average 
over many realizations. The results confirm that during the spreading 
dynamics it is very likely that highly connected nodes are reached 
by the update. In fact $ 1-R_k$ decreases exponentially with $k$ and very 
high levels of reliability  are obtained well before 
the natural cut-off of the  network ($k_{max}\sim 10^2$) even 
for moderate values of $\alpha$. This is an interesting feature
signaling that heterogeneous topologies can be considered reliable as
far as the "hubs" are concerned. We should note, however, that since the
reliability is not a model independent quantity,  this result 
might depend upon the specific details of the rumor algorithm.

\begin{figure}[t]
\begin{center}
\epsfig{file=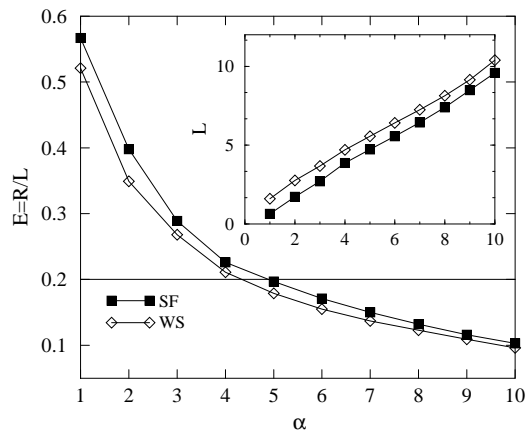,width=2.3in,clip=1,angle=-90}
\end{center}
\caption{Efficiency of the rumor spreading process as a function of
$\alpha$ in networks of size $N=10^4$. Results are compared with 
the efficiency of the basic broadcast 
algorithm that is represented by the thick line. The 
inset shows the growth of the load generated as a function of the 
parameter $\alpha$.}
\label{figure2}
\end{figure}

In general, one does not only want to have high reliability levels,
but also the lowest cost in terms of network load \cite{ep1,ep3}.
This is generally achieved by imposing the minimum possible load to
the network.  Here, we define the load $L$ imposed to the network as
the number of contacts established per node, i.e., how many messages
on average each node sends to its neighbors in order to propagate the
update. By using this quantity, an obvious definition of the global
efficiency $E$ of the whole process is represented by the number of
individuals who have got the update per unit of load,
$E=\frac{R}{L}$. Its physical meaning is straightforward: the
efficiency is equal to the fraction of ``useful messages'' (number of
sites reached by the rumor) over total ``load'' imposed to the system.
In Fig.~\ref{figure2}, we report the behavior of the efficiency of the
spreading process as a function of $\alpha$. In this case, the
scale-free topology appears the most efficient for data
dissemination. Indeed, the relative difference between SF and WS
networks in global efficiency is larger than $10\%$ up to values of
$\alpha=5$.  This can be appreciated also by looking at the inset in
Fig.~\ref{figure2}. For both topologies the load on the network grows
with $\alpha$, but the load imposed on SF networks is always smaller
than on WS nets. Finally, it is interesting to compare the epidemic
algorithm efficiency with those of the simplest {\em broadcast}
strategy. This strategy essentially consists of a deterministic
message-passing of each element to all its neighbors except the one
from which the first update has been received.  This way, a
reliability $R=1$ is achieved since all nodes are surely contacted. In
this case, the load is simply given by $\langle k\rangle-1$. In the
case of networks with $\langle k \rangle=6$ as the one used in the
present study, the efficiency of the broadcast strategy is therefore
$E=0.2$. It is interesting to note that in both the BA and the WS
case, the epidemic algorithm achieve a better efficiency for a wide
range of values of the parameter $\alpha$. This indicates that
epidemic algorithms can provide attractive alternatives to broadcast
solutions so far as the efficiency is concerned.

It is worth stressing that we are considering strategies in which the
nodes do not have memory; i.e. they may try to resend a message to a
node that has been contacted before. It is possible, however, to
conceive different dynamics of the updating spreading strategies in
which a trade-off between the memory introduced in the process and the
optimization of reliability and efficiency is opportunely chosen.
Other options rely on a careful tuning of the local message sending
dynamics.  For instance, we considered the case in which each spreader
contacts only one node at each time step, reducing the effects of the
hubs.  This simple change allows for higher levels of efficiency,
however, at the price of a much lower reliability.  In general, thus
it is possible to devise and tailor different processes that optimize
one or more features of the update spreading on a given topology.  We
defer a more detailed study of this issue to future work.

In summary, we have studied the effect of the complex topological
properties of many real networks in  epidemic strategies for the
communication of updates. The obtained results 
stimulate the seeking of heuristics and analytical 
methods to optimize epidemic algorithms taking into account the 
specific topology of the underlying network. These studies may have a
large impact in  technological and
communication networks where the use of rumor-like algorithms might
become a practice for data dissemination, reliable group
communication or replicated database maintenance.  Finally,
they might provide a deeper understanding for social phenomena such as
the spreading of new ideas in a population or the efficiency of
marketing campaigns.

\begin{acknowledgments}
We are grateful to A. Barrat and R. Pastor-Satorras for helpful
comments and suggestions.  Y.\ M.\ acknowledges financial support from
the Secretar\'{\i}a de Estado de Educaci\'on y Universidades (Spain,
SB2000-0357) and the hospitality of the Complexity Research Group of
BT Exact, U.K. A.\ V.\ is supported by the EC - Fet Open project COSIN
IST-2001-33555. This work has been partially supported by the Spanish
DGICYT project BFM2002-01798.
\end{acknowledgments}


\begin{thebibliography}{99}
\bibitem{note1} The model proposed here has been considered 
before keeping $\lambda=\alpha=1$~\cite{zanette} in Watts-Srogatz 
networks with different level of radnomness in the connectivity pattern.
\end{thebibliography}
\end{document}